Local Studies of the Ferromagnetic Ordering Temperature Suppression in SrRuO$_3$*

Z. H. Han, J. I. Budnick, M. Daniel, W. A. Hines, and
D. M. Pease

Department of Physics and Institute of Materials Science,
University of Connecticut, Storrs, CT 06269, USA

AND

P. W. Klamut, B. Dabrowski, S. M. Mini, M. Maxwell, and
C. W. Kimball

Department of Physics, Northern Illinois University, DeKalb,
IL 60115, USA

In order to gain insight into suppression of the ferromagnetic ordering temperature, local studies utilizing $^{99,101}$Ru zero-field spin-echo NMR, along with complimentary magnetization, high-angle x-ray diffraction, and Ru K-edge XAFS measurements, have been carried out on samples of SrRuO$_3$ annealed at both "ambient" (atmospheric) pressure and "high-pressure" oxygen (600 atm). For the "high-pressure" sample, the results indicate structural disorder and the existence of some vacancies on the Ru sites, along with a reduced magnetic ordering temperature.

## 1. Introduction

Among the ruthenium oxides, strontium ruthenate, SrRuO$_3$, which has a distorted perovskite structure, is the only known 4d transition-metal oxide which is ferromagnetic. Previous studies have reported a suppression of the ferromagnetic ordering temperature for SrRuO$_3$ due to: (1) the substitution of Ca and Na [1,2], (2) the effects of strain occurring in thin films [3], and (3) the synthesis conditions [4,5]. In particular, the synthesis of SrRuO$_3$ under high-pressure oxygen produces a nonstoichiometric form with randomly distributed vacancies on the Ru-sites, along with a significantly reduced ferromagnetic ordering temperature [5]. The decrease in the ordering temperature was attributed to an increase in the Ru formal valence along with structural disorder. In order to understand the microscopic origin of the

---

suppression of ferromagnetism in SrRuO$_3$, local studies utilizing $^{99,101}$Ru zero-field spin-echo nuclear magnetic resonance (NMR), along with complimentary magnetization, high-angle x-ray diffraction (XRD), and Ru K-edge x-ray absorption fine stucture (XAFS), have been carried out and are reported here.

## 2. Experimental Procedure, Results, and Discussion

Polycrystalline SrRuO$_3$ was synthesized from a stoichiometric mixture of SrCO$_3$ and RuO$_2$ using a solid state reaction method [5]. The parent SrRuO$_3$ material was subdivided and processsed under the following conditions: (1) annealed twice under flowing oxygen at 1050 °C for 10 h ("ambient") and (2) annealed twice under 600 atm oxygen at 1050 °C for 10 h ("high-pressure"). In this work, detailed zero-field spin-echo $^{99,101}$Ru NMR spectra were obtained at 1.3 K and 4.2 K over the frequency range from 45 to 105 MHz using a phase-coherent spectrometer. In addition, magnetization measurements using a SQUID magnetometer and XRD analysis using a diffractometer with Cu K$\alpha$I radiation were carried out. The XAFS experiments were carried out in the transmission mode at the X-11A beamline of the National Syncrotron Light Source, Brookhaven National Laboratory.

Figure 1 shows the zero-field spin-echo NMR spectra obtained for the "ambient" sample at 4.2 K (open symbols) and high-pressure" sample at 1.3 K (closed symbols), respectively. Consistent with earlier work, the NMR spectrum for "ambient" SrRuO$_3$ consists of two well-defined peaks at 64.4 MHz and 72.2 MHz corresponding to the $^{99}$Ru and $^{101}$Ru isotopes, respectively, and a hyperfine field of 329 kG [6]. For the "high-pressure" sample, although the magnetization measurements show a lower ferromagnetic ordering temperature (90 K compared to 160 K for the "ambient" sample), the NMR spectrum still shows the same two peaks at 64.4 MHz and 72.2 MHz, attributed again to $^{99}$Ru and $^{101}$Ru, respectively, i.e., there is no significant shift. However, the two peaks exhibit considerable broadening, along with structure on both the low and high frequency sides which is believed to be quadrupolar in origin. These peak frequencies and the corresponding hyperfine field value of 329 kG are consistent with the Ru$^{+4}$ or low-spin (S = 1) valence state. The $^{99,101}$Ru NMR spectrum is qualitatively similar to those observed for the Ru sites having the Ru$^{+4}$ valence state in both RuSr$_2$GdCu$_2$O$_8$ [7] and RuSr$_2$YCu$_2$O$_8$ [8]. The spectra are characterized by a large internal hyperfine field perturbed by a smaller quadrupole interaction; the Ru sites are located in distorted RuO$_6$ octahedra. The observed quadrupole interactions are large enough to account for the broadening shown in Fig. 1 for the "high-pressure" sample. Furthermore, the effect of the high-pressure oxygen treatment is much more severe on the

$^{101}$Ru peak, which is consistent with the fact that the quadrupole moment for $^{101}$Ru is almost six times larger than that for $^{99}$Ru. A search over the frequency range 100 MHz to 150 MHz yielded no evidence for the existence of the Ru$^{+5}$ or high-spin (S = 3/2) valence state. Measurements of the (homogeneous) spin-spin relaxation time $T_2$ were made at both peaks for the two samples. For the "ambient" sample, the relaxation behavior could be described by a single exponential with $T_2 = 1,100$ $\mu$s and 920 $\mu$s for the $^{99}$Ru and $^{101}$Ru peaks, respectively. On the other hand, the relaxation behavior for the "high-pressure" sample could not be described by a single exponential; however, the relaxation rates were significantly faster. Finally, it should be noted that NMR spin-echo signal was considerably weaker for the "high-pressure" sample compared to that for the "ambient" sample, which required operation at 1.3 K. Of course, the severe broadening accounts for a large part of the signal reduction; however, the NMR enhancement factor appeared to be reduced as well for the "high-pressure" sample. This is consistent with the increased coercive field observed for the "high-pressure" sample and suggests an enhanced pinning of the domain walls.

For both the "ambient" and "high-pressure" samples, a direct measure of the magnetic ordering temperature $T_c$ was made by observing the zero-field-cooled and field-cooled dc magnetization in a field of 50 G. The values of $T_c = 160$ K and 90 K obtained for the "ambient" and "high pressure" samples, respectively, are consistent with earlier results using both dc magnetization and ac susceptiblity measurements [5]. The low temperature magnetic moment per Ru atom did not saturate; values of 1.25 $\mu_B$ and 0.86 $\mu_B$, were obtained at 50 kG for the "ambient" and "high-pressure" samples, respectively. These values assume complete occupancy of the Ru sites. Curie-Weiss fits to the magnetic susceptibility in the paramagnetic state were made for both samples. The calculated paramagnetic moment (Curie temperature) values were 2.66 $\mu_B$ (153 K) and 2.61 $\mu_B$ (97 K) for the "ambient" and "high-pressure" samples, respectively, assuming complete occupancy of the Ru sites. Although the uncertainty involved in the Curie-Weiss fit is comparable to the difference in the moment values, the lower value for the "high pressure" oxygen sample is consistent with some vacancies at the Ru sites. The temperature dependence of the magnetization measured at 10 kG could be fit to the Bloch $T^{3/2}$-law up to approximately 0.75 $T_c$ for both samples. Measurements of the complete hysteresis loop for both the "ambient" and "high-pressure" samples, indicated a considerably larger value of the coercive field for the latter. This suggests an enhanced pinning of the domain walls in the "high-pressure" sample. The "ambient" and "high-pressure" SrRuO$_3$ samples are characterized by the GdFeO$_3$-like orthorhombic structure. The lattice parameters a, b, and c were the same for the two samples within the accuracy of the measurement. Furthermore, in an attempt to observe any small subtle differences between the XRD patterns for the "high-pressure" and "ambient" samples, a detailed scan was made of two high-angle lines: (1) near 67° which consists of the (400) and (224) reflections and (2) near 77° which consists of the (116) reflection. Within the accuracy of the measurement, the scans for the two were identical. Finally, preliminary results from the Ru K-edge XAFS measurements indicate the existence of some disorder in the Ru-O bond for the "high-pressure" sample.

Work at UC was supported by the NSF (DMR-9705136), and at NIU by the NSF (DMR-0105398) and by the State of Illinois under HECA.

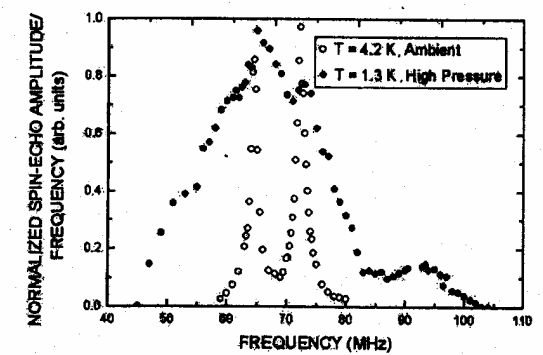

Fig. 1. $^{99,101}$Ru zero-field spin-echo NMR spectra for SrRuO$_3$